\documentclass[epj]{svjour}

\usepackage[dvips]{graphicx}

\usepackage{array}
\usepackage{amssymb}
\usepackage{amsmath}
\usepackage{hhline}
\usepackage{longtable}
\usepackage{dcolumn}
\usepackage{bm}
\usepackage{subfigure}
\usepackage{epsfig}
\usepackage{latexsym,amsmath}
\usepackage{amsbsy}

\makeatletter
\renewcommand\appendix{\par
  \setcounter{section}{0}%
  \setcounter{subsection}{0}%
  \setcounter{equation}{0}%
  \renewcommand\theequation{\Alph{section}.\arabic{equation}}
  \renewcommand\thesection{Appendix \@Alph\c@section:}}
\makeatother

\begin{document}
\title{Role of fractal dimension in random walks on scale-free networks}

\author{Zhongzhi Zhang \thanks{e-mail: zhangzz@fudan.edu.cn}  \and Yihang Yang \and Shuyang Gao}                     
\institute{School of Computer Science, Fudan University, Shanghai
200433, China \and Shanghai Key Lab of Intelligent Information
Processing, Fudan University, Shanghai 200433, China}

\date{Received: date / Revised version: date}

\abstract{Fractal dimension is central to understanding dynamical processes occurring on networks; however, the relation between fractal dimension and random walks on fractal scale-free networks has been rarely addressed, despite the fact that such networks are ubiquitous in real-life world. In this paper, we study the trapping problem on two families of networks. The first is deterministic, often called $(x,y)$-flowers; the other is random, which is a combination of $(1,3)$-flower and $(2,4)$-flower and thus called hybrid networks. The two network families display rich behavior as observed in various real systems, as well as some unique topological properties not shared by other networks. We derive analytically the average trapping time for random walks on both the $(x,y)$-flowers and the hybrid networks with an immobile trap positioned at an initial node, i.e., a hub node with the highest degree in the networks. Based on these analytical formulae, we show how the average trapping time scales with the network size. Comparing the obtained results, we further uncover that fractal dimension plays a decisive role in the behavior of average trapping time on fractal scale-free networks, i.e., the average trapping time decreases with an increasing fractal dimension.
\PACS{{05.40.Fb}{Random walks and Levy flights}   \and
      {89.75.Hc}{Networks and genealogical trees}   \and
      {05.60.Cd}{Classical transport} \and
      {89.75.Da}{Systems obeying scaling laws}
      } 
} 

 \maketitle

\section{introduction}

Complex networks have become a powerful and common tool for studying complex networked systems in nature and society, which allow for describing quantitatively the structural features and complexity of real systems~\cite{AlBa02,Ne03}. A fundamental problem in the field of complex networks is to unveil the influence of topology on diverse dynamical processes taking places on networks~\cite{DoGoMe08}. As an important stochastic process, random walks are a subject of a large volume of research~\cite{HaBe87,MeKl00,MeKl04,BuCa05,BeLoMoVo11}, due to their wide range of applications in various areas~\cite{Sp1964,We1994,Hu1996}. A primarily interesting quantity related to random walks is the first-passage time (FPT)~\cite{Re01,NoRi04,Bobe05,CoBeTeVoKl07,BeMeTeVo08}. Recently, the mean first-passage time (MFPT) to a target node on a network has received an increasing attention~\cite{Mo69,KaBa02PRE,KaBa02IJBC,Ag08,CaAb08,HaRo08,BeTuKo10,TeBeVo11}, since it can serve as a measure of search efficiency to find the target.

A remarkable properties unveiled by extensive empirical research is that many, perhaps most real-life networks display scale-free phenomenon, meaning that their degree distribution $P(k)$ follows a power law $P(k) \sim k^{-\gamma}$~\cite{BaAl99}. This property constitutes our basic understanding of the structural organization of real systems and has a profound impact on random walks occurring on scale-free networks. A lot of work showed that search efficiency is substantially improved if the target is a node with the highest degree in scale-free networks~\cite{KiCaHaAr08,ZhQiZhXiGu09,ZhGuXiQiZh09,ZhZhXiChLiGu09,AgBu09,TeBeVo09,ZhGaXi10,AgBuMa10}. Nevertheless, scale-free behavior cannot reflect all the structural information of real networks. It was acknowledged that in addition the power-law property, a variety of real-life systems exhibit fractal scaling, which is often characterized by the fractal dimension~\cite{SoHaMa05}. Taking into account fractal scaling of scale-free networks can lead to a better understanding of how the underlying systems work~\cite{SoHaMa06}. It was shown that fractality is unfavorable for the target search problem in fractal scale-free networks~\cite{ZhXiZhGaGu09,ZhLiMa11}. However, what is the relation between the fractal dimension and scaling of MFPT for random walks on scale-free networks remains not well understood.

In this paper, we study the classic trapping problem on two families of networks, which is a random-walk issue with an immobile trap located at a fixed position absorbing all walkers that visit it. The first network family is of a deterministic type, called $(x,y)$-flowers~\cite{RoHaAv07,RoAv07} that contains the intensively studied pseudofractal scale-free web~\cite{DoGoMe02,ZhZhCh07} and fractal hierarchical lattices~\cite{BeOs79,KaGr81,GrKa82,HiBe06} as their limiting cases by adjusting the parameters $x$ and $y$. The second family is random that can be looked on as a mixture of $(1,3)$-flower and $(2,4)$-flower and thus called hybrid networks~\cite{ZhZhZoChGu09}. Both network families present some typical topological properties as observed in real-life networks. We focus on a particular case of the trapping problem where the trap is fixed at an initial node (namely, node with the largest degree). We derive analytically the average trapping time for both the $(x,y)$-flowers and the hybrid networks, based on which we obtain the dependence relation of average trapping time on the network size. Our results show that in both network families, the average trapping time grows algebraically with their size. Lastly, we present a detailed analysis of the obtained results, and show that the increasing fractal dimensions of fractal  scale-free networks may lead to enhancement of search efficiency for those target nodes with the highest degree.

\section{\label{SecModel}Network models and their properties}

Here we introduce two families of networks and their structural features. The first one is deterministic, while the second one is stochastic.

\subsection{\label{SecModelA}Construction and topology of $(x,y)$-flowers}

\begin{figure}[h]
\begin{center}
\includegraphics[width=0.65\linewidth,trim=50 0 50 10]{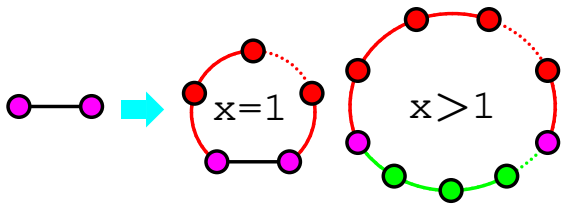}
\caption{(Color online) Construction approach of the $(x,y)$-flowers. To produce the networks at iteration $n+1$, we can replace each edge in the networks at iteration $n$ by two parallel paths with lengths $x$ ($x\geq 1$) and $y$ ($y\geq x$ and $y>
1$) on the right-hand side of the arrow. When $x=1$, any pair of old nodes directly linked to each other by an old edge creates $y-1$ new nodes. These $y-1$ new nodes and the two old ones form a red path of length $y$; while the old edge is remained. When $x>1$, each old edge connecting two old nodes is deleted and simultaneously replaced by two paths with the two old nodes as the endpoints of both paths: the $y-1$ red nodes and the two old nodes form a path of length $y$, while the $x-1$ green nodes and the two old nodes form the other path with length $x$. }\label{fig1}
\end{center}
\end{figure}

\begin{figure}
\begin{center}
\includegraphics[width=0.60\linewidth,trim=0 0 0 0]{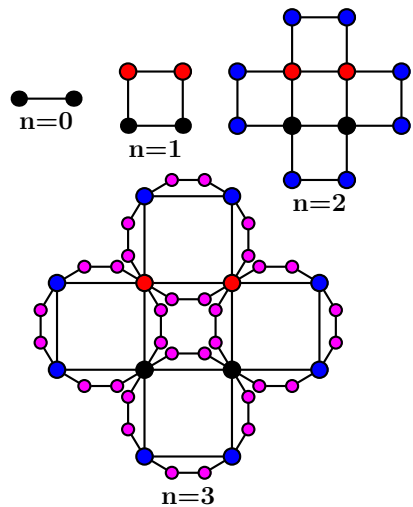}
\caption{(Color online) Illustration for the evolution process of the $(1,3)$-flower.} \label{flower}
\end{center}
\end{figure}

\begin{figure}[h]
\centering
\includegraphics[width=0.7\linewidth,trim=0 0 0 0]{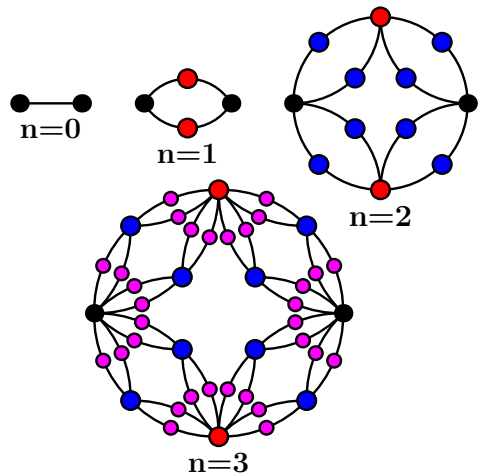}
\caption{(Color online) Iterative growth for the $(2,2)$-flower.}\label{fractal}
\end{figure}

The first family of studied networks, frequently called $(x,y)$-flowers~\cite{RoHaAv07,RoAv07}, are constructed in an iterative way. Let $F_{n}(x,y)$ ($n \geq 0$) denote the $(x,y)$-flowers after $n$ iterations. Without loss of generality we assume that $x \le y$ and $y>1$. The $(x,y)$-flowers are built in the following way, see Fig.~\ref{fig1}. For $n=0$, $F_{0}(x,y)$ consists of two initial nodes connected by an edge. For $n \ge 1$, $F_{n}(x,y)$ is generated from $F_{n-1}(x,y)$ through replacing each exiting edge in $F_{n-1}(x,y)$ by two parallel paths consisting of $x$ and $y$ links. Figures~\ref{flower} and~\ref{fractal} illustrate the iterative processes of two special networks: $(1,3)$-flower and $(2,2)$-flower.

The iterative construction allows for determining relevant properties of the $(x,y)$-flowers. It is easy to derive that the numbers of nodes and edges in $F_{n}(x,y)$ are
\begin{equation}\label{Nn}
N_{n}=\frac{x+y-2}{x+y-1}(x+y)^{n}+\frac{x+y}{x+y-1}\,
\end{equation}
and
\begin{equation}\label{Mn}
M_{n} = (x+y)^n\,,
\end{equation}
respectively. And the average degree of $F_{n}(x,y)$ is
\begin{equation}\label{kn}
\langle k \rangle _{n} = \frac{2M_{n}}{N_{n}}=\frac{2(x+y-1)(x+y)^{n-1}}{(x+y-2)(x+y)^{n-1}+1}\,,
\end{equation}
which approaches $2(x+y-1)/(x+y-2)$ as $n\rightarrow \infty$.

The $(x,y)$-flowers display rich behavior in their topological structure~\cite{RoHaAv07,RoAv07}. They follow a power-law degree distribution $P(k) \sim k^{-\gamma}$ with the exponent $\gamma = 1 +\ln(x+y)/\ln2$ belonging to the interval $[1+\ln 3/ \ln2,\infty)$. For $x=1$, the $(x,y)$-flowers are small-world~\cite{WaSt98} but non-fractal; while for $x>1$, the networks are ``large-world" and fractal with the fractal dimension $d_{\rm f}=\ln(x+y)/\ln x$. In addition, in the whole family of $F_{n}(x,y)$, all networks with the same parameter $x+y$ have the same identical degree sequence, hence the same degree distribution.

\subsection{\label{SecModelB}Generation and architecture of hybrid networks}

\begin{figure}[h]
\begin{center}
\includegraphics[width=0.6\linewidth,trim=100 0 100 10]{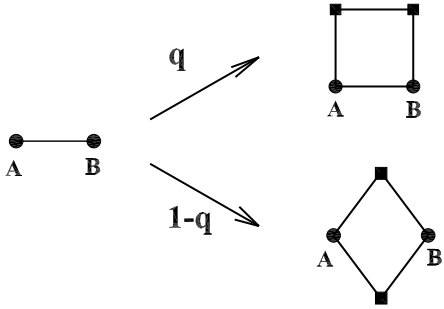}
\caption{Iterative construction method of the hybrid networks. Each edge is
replaced by either of the connected clusters on the right-hand side
of arrows with a certain probability, where black squares represent
new nodes.}\label{built}
\end{center}
\end{figure}

Different from the first family of networks, the second family of networks in question are built iteratively but randomly~\cite{ZhZhZoChGu09},
see Fig.~\ref{built}. Since the networks are a random mixture of $(1,3)$-flower and $(2,2)$-flower, we call them hybrid networks. Let $H_{n}$ ($n \geq 0$) represent the
hybrid networks after $n$ iterations. Then the networks are constructed as
follows. For $n=0$, the network $H_{0}$ is composed of two initial
nodes connected by an edge. For $n \geq 1$,
$H_{n}$ is derived from $H_{n-1}$: replace each link existing in $H_{n-1}$ either by a
square on the top right of Fig.~\ref{built} with
probability $q$, or by a diamond on the bottom right
with probability $1-q$. For $q=1$ and $q=0$, both networks are deterministic, which are separately $(1,3)$-flower and $(2,2)$-flower, see Figs.~\ref{flower} and~\ref{fractal}.

The number of edges and nodes in $H_n$ are $M_n =4^n$
$N_n = 2(4^n+2)/3$, respectively. The average degree of $H_n$ is
$\langle k \rangle_n = 2M_n/N_n=(3\times 4^n)/(4^n+2)$, which approaches
3 in the limit of large $n$. In the full range of $0\leq q \leq 1$, the whole family of networks has identical degree sequence and degree distribution with a power-law form $P(k) \sim k^{-3}$. The clustering coefficient of the networks is zero, since they
contain no triangles. For $q=0$, the network is ``large-world" and fractal with a fractal dimension $d_{\rm f}=2$, while for $q=1$ the network is small-world and non-fractal with an infinite fractal dimension. Thus, when parameter $q$ increases from 0 to 1, the
network family exhibits a crossover from ``large" to small worlds, and simultaneously undergoes a transition from fractal to non-fractal scaling, i.e., its fractal dimension grows from 2 to $\infty$.

The peculiar structure of the two network families makes it possible to explore the role of fractal dimension, instead of the power-law degree distribution, in the behavior of random walks on scale-free networks, since some particular cases of them share the same degree distribution. This is the main goal of this paper.

\section{Random walks with a trap fixed on an initial node}

Having introducing the networks, we investigate the simple discrete-time random
walks on $F_{n}(x,y)$ and $H_{n}$. At each time step, the
walker moves uniformly from its current location to one of its
neighboring nodes. Here we concentrate on a special random-walk issue with a trap
placed at an initial node denoted by $i_T$, which is also known as the
trapping problem. One of the most significant quantity
characterizing the trapping problem is the trapping time, i.e., the so-called first-passage time, which is defined as the
expected time spent by a walker, starting from a source node, to arrive at the trap for the first
time. The average trapping time, namely, mean first-passage time to the trap node, defined as the average of trapping time over all starting points other than the trap, measures the efficiency of the trapping process: the less the average trapping time, the higher the efficiency,
and vice versa.

Let $T_i^{(n)}$ stand for the trapping time for a walker starting off from node $i$
to first visit the trap $i_T$ in $F_{n}(x,y)$ or $H_{n}$.
Then, the average trapping time, $\langle T \rangle_n$, which is
the mean of $T_i^{(n)}$ over all non-trap starting nodes in $F_{n}(x,y)$ or $H_{n}$, is given by
\begin{equation}\label{ATT}
 \langle T
\rangle_n=\frac{1}{N_n-1}\sum_{i\in \Omega_n}T_i^{(n)}\,,
\end{equation}
where $\Omega_n$ denote the set of nodes in $F_{n}(x,y)$ or $H_{n}$. It is obvious that for all $n$ we have $T_{i_T}^{(n)}=0$.
In the following, we will find the exact
solution to $\langle T \rangle_n$ for both $F_{n}(x,y)$ or $H_{n}$, as well as the dependence
relation of $\langle T \rangle_n$ on the network size $N_n$.

\subsection{Average trapping time for $(x,y)$-flowers}

We study how the trapping time $T_i^{(n)}$ changes
with $n$. 
According to the construction algorithm, for an edge incident an old node $i$ and one of its old neighbors $j$ at iteration $n$, it will generate $x+y-2$ new nodes at iteration $n+1$, among which $x-1$ nodes, denoted by $v_1$,
$v_2$, $\ldots$, $v_{x-1}$, together with nodes $i$ and $j$ form a path ($i,v_1,
v_2, \ldots, v_{x-1},j$) of $x$ links, and the other $y-1$ nodes, denoted by $u_1$, $u_2$, $\ldots$, $u_{y-1}$, together with $i$ and $j$ constitute a path ($i,u_1,
u_2, \ldots, u_{y-1},j$) of $y$ links. We call $v_1$ and $u_1$
first-order (direct) neighbors of $i$, $v_2$ and $u_2$ second-order
neighbor of $i$, $v_3$ and $u_3$ third-order neighbor of $i$, and so on.

Let's examine a random walk on $F_{n+1}(x,y)$. By construction, upon
growth of the networks from generation $n$ to next generation $n+1$, the degree of
node $i$ doubles, that is, it grows from $k_i(n)$ to $k_i(n+1)=2\,k_i(n)$.
Among these $2\,k_i(n)$ neighbors,
one half are the first-order neighbors of $i$ with each belonging to a path of $x$ links, and the other half are the first-order neighbors of $i$, that belong to the paths of $y$ links. Let $Z$ be the FPT going from node $i$ to any of its $k_i(n)$ old
neighbors, i.e., those nodes directly connected to node $i$ at iteration $n$;
let $X_l$ ($1 \leq l \leq x-1$) be the FPT originating at
any of $i'$s $k_i(n)$ $l$th order new neighbors belonging to a path of $x$ links to one of $i'$s $k_i(n)$ old neighbors; and
let $Y_m$ ($1 \leq m \leq y-1$) be the FPT for going from
any of the $k_i(n)$ $m$th order new neighbors of $i$, belonging to a path of $y$ links, to one of $i'$s $k_i(n)$ old neighbors. These FPTs obey the following backward
equations:
\begin{eqnarray}\label{MFPT6}
\left\{
\begin{array}{ccc}
Z&=&\frac{1}{2}(1+X_1) + \frac{1}{2}(1+Y_1)\,,\\
X_1&=&\frac{1}{2}(1+Z) + \frac{1}{2}(1+X_2)\,,\\
X_2&=&\frac{1}{2}(1+X_1) + \frac{1}{2}(1+X_3)\,,\\
\vdots&\quad& \quad  \quad \vdots \quad\quad \quad\quad \vdots \quad\quad \\
X_{x-1}&=&\frac{1}{2}+ \frac{1}{2}(1+X_{x-2})\,,\\
Y_1&=&\frac{1}{2}(1+Z) + \frac{1}{2}(1+Y_2)\,,\\
Y_2&=&\frac{1}{2}(1+Y_1) + \frac{1}{2}(1+Y_3)\,,\\
\vdots&\quad& \quad  \quad \vdots \quad\quad \quad\quad \vdots \quad\quad \\
Y_{y-1}&=&\frac{1}{2}+ \frac{1}{2}(1+Y_{y-2})\,.
 \end{array}
 \right.
\end{eqnarray}

Equation~(\ref{MFPT6}) has a solution $Z=xy$. Thus, upon the growth of the $(x,y)-$flowers from generation $n$ to generation $n+1$, the
FPT from an arbitrary node $i$ to another node $j$ (both $i$ and
$j$ are already existing in $F_{n}(x,y)$) increases by a factor of $xy$. Then, we
have $T_i^{(n+1)}=xy\,T_i^{(n)}$, which is a basic feature of random walks taking place on the $(x,y)$-flowers and will be very important for the following derivation
of the exact formula for the average trapping time.

In order to determine the average trapping time $\langle T \rangle_n$, one may alternatively evaluate $\sum_{i\in \Omega_n}T_i^{(n)}$. To this end, we introduce some
variables that will be convenient for description. Let
$\bar{\Omega}_n$ denote the set of all nodes in $F_n(x,y)$ that enter the networks at iteration $n$. Then, we can further define the following intermediate quantities for $1\leq g \leq n$:
\begin{equation}
T_{g, {\rm tot}}^{(n)} = \sum_{i \in \Omega_g} T_i^{(n)},
\end{equation}
and
\begin{equation}
\bar{T}_{g, {\rm tot}}^{(n)} = \sum_{i \in \bar{\Omega}_g} T_i^{(n)}.
\end{equation}
Hence, the problem of determining $\langle T \rangle_n$ is reduced to finding $T_{n,{\rm tot}}^{(n)}$. It is evident that $\Omega_n = \bar{\Omega}_n \cup
\Omega_{n-1}$, which leads to
\begin{equation}\label{eq:Ttotxy}
T_{n, {\rm tot}}^{(n)} = T_{n - 1, {\rm tot}}^{(n)}+\bar{T}_{n, {\rm tot}}^{(n)}= xy\,T_{n - 1, {\rm tot}}^{(n-1)} + \bar{T}_{n, {\rm tot}}^{(n)}\,,
\end{equation}
where we have used the relation $T_i^{(n)}=xy\,T_i^{(n-1)}$. Equation~(\ref{eq:Ttotxy}) indicates that before calculating $T_{n, {\rm
tot}}^{(n)}$, we should first evaluating the quantity $\bar{T}_{n,
{\rm tot}}^{(n)}$ that can be derived as follows.

As mentioned above, for each edge attaching $i$ and $j$ in $F_n(x,y)$, it will generate $x+y-2$ new
nodes at iteration $n+1$, say $v_1$, $v_2$, $\ldots$, $v_{x-1}$, $u_1$, $u_2$, $\ldots$, $u_{y-1}$. For a random walk in $F_{n+1}(x,y)$, the trapping times of these $x+y-2$ new nodes obey the following relations:
\begin{eqnarray}\label{MFPT6xy}
\left\{
\begin{array}{ccc}

T_{v_1}^{(n+1)} &=& \frac{1}{2}\left(1+T_{i}^{(n+1)}\right)+\frac{1}{2}\left(1+T_{v_2}^{(n+1)}\right)\,,\\
T_{v_2}^{(n+1)} &=& \frac{1}{2}\left(1+T_{v_1}^{(n+1)}\right)+\frac{1}{2}\left(1+T_{v_3}^{(n+1)}\right)\,,\\
\vdots&\quad& \quad  \quad \vdots \quad\quad \quad\quad \vdots \quad\quad \\
T_{v_{x-1}}^{(n+1)} &=& \frac{1}{2}\left(1+T_{v_{x-2}}^{(n+1)})\right)+\frac{1}{2}\left(1+T_{j}^{(n+1)}\right)\,,\\

T_{u_1}^{(n+1)} &=& \frac{1}{2}\left(1+T_{i}^{(n+1)}\right)+\frac{1}{2}\left(1+T_{u_2}^{(n+1)}\right)\,,\\
T_{u_2}^{(n+1)} &=& \frac{1}{2}\left(1+T_{u_1}^{(n+1)}\right)+\frac{1}{2}\left(1+T_{u_3}^{(n+1)}\right)\,,\\
\vdots&\quad& \quad  \quad \vdots \quad\quad \quad\quad \vdots \quad\quad \\
T_{u_{y-1}}^{(n+1)} &=& \frac{1}{2}\left(1+T_{u_{y-2}}^{(n+1)})\right)+\frac{1}{2}\left(1+T_{j}^{(n+1)}\right)\,.

\end{array}
\right.
\end{eqnarray}

From Eq.~(\ref{MFPT6xy}), we have
\begin{eqnarray}
T_{v_1}^{(n+1)} +T_{v_{x-1}}^{(n+1)}= 2(x-1) + T_{i}^{(n+1)} + T_{j}^{(n+1)}\,,\nonumber \\
T_{v_2}^{(n+1)} +T_{v_{x-2}}^{(n+1)} = 2(x-3) +T_{v_1}^{(n+1)} +T_{v_{x-1}}^{(n+1)}\,, \nonumber \\
T_{v_3}^{(n+1)} +T_{v_{x-3}}^{(n+1)} = 2(x-5) + T_{v_2}^{(n+1)} +T_{v_{x-2}}^{(n+1)} \,,\nonumber
\end{eqnarray}
and so on. From these relations, we obtain
\begin{equation}\label{eq:xyv}
\sum_{l=1}^{x - 1} T_{v_l}^{(n+1)} = \frac{x(x^2-1)}{6} + \frac{x -
1}{2}\left(T_{i}^{(n+1)} + T_{j}^{(n+1)}\right)\,.
\end{equation}
In a similar way, we can derive that
\begin{equation}\label{eq:xyu}
\sum_{m=1}^{y - 1} T_{u_m}^{(n+1)} = \frac{y(y^2-1)}{6} + \frac{y -
1}{2}\left(T_{i}^{(n+1)} + T_{j}^{(n+1)}\right)\,.
\end{equation}

Summing Eqs.~(\ref{eq:xyv}) and~(\ref{eq:xyu}) over all the $M_n$ edges
pre-existing in $F_n(x,y)$ leads to
\begin{eqnarray}\label{eq:rec1xy}
\bar{T}_{n+1, {\rm tot}}^{(n+1)} &=& \frac{x(x^2-1)+y(y^2-1)}{6}\,M_n \nonumber\\ &\quad&
+ \sum_{i \in \Omega_{n}}\left(k_i(n)\times \frac{x+y-2}{2} T_i^{(n+1)}\right) \nonumber\\
&=& \frac{x(x^2-1)+y(y^2-1)}{6}(x+y)^n + \nonumber\\ &\quad&
(x+y-2)\,\bar{T}_{n, {\rm tot}}^{(n+1)} + 2 \, (x+y-2)\,\bar{T}_{n-1, {\rm tot}}^{(n+1)} \nonumber\\ &\quad&
+ \cdots+2^{n-1}\, (x+y-2) \,\bar{T}_{1, {\rm tot}}^{(n+1)}\nonumber + \\
&\quad& 2^{n-1} \, (x+y-2)\,\bar{T}_{0, {\rm tot}}^{(n+1)}\,.
\end{eqnarray}
Analogously, we have
\begin{eqnarray}\label{eq:rec2xy}
\bar{T}_{n+2, {\rm tot}}^{(n+2)}
&=& \frac{x(x^2-1)+y(y^2-1)}{6}(x+y)^{n+1} + \nonumber\\ &\quad&
(x+y-2)\,\bar{T}_{n+1, {\rm tot}}^{(n+2)} + 2 \, (x+y-2)\,\bar{T}_{n, {\rm tot}}^{(n+2)} \nonumber\\ &\quad&
+ \cdots+2^{n}\, (x+y-2) \,\bar{T}_{1, {\rm tot}}^{(n+2)}\nonumber + \\
&\quad& 2^{n} \, (x+y-2)\,\bar{T}_{0, {\rm tot}}^{(n+2)}\,.
\end{eqnarray}

Equation~(\ref{eq:rec2xy}) minus Eq.~(\ref{eq:rec1xy}) times $2xy$
and considering the relation $T_i^{(n+2)}=xy \,T_i^{(n+1)}$, we obtain the
recursion relation:
\begin{equation} \label{eq:T1xy}
\begin{aligned}
\bar{T}_{n+2,{\rm tot}}^{(n+2)}=
&xy(x+y)\,\bar{T}_{n+1, {\rm tot}}^{(n+1)}+\\
&\frac{x(x^2-1)+y(y^2-1)}{6}(x+y-2xy)(x+y)^n\,.
\end{aligned}
\end{equation}
Using the initial condition $\bar{T}_{1, {\rm tot}}^{(1)} = \frac{(x+y)^3-(x+y)}{6}-xy$ Eq.~(\ref{eq:T1xy}) is solved inductively to obtain
\begin{equation} \label{eq:TXxy}
\bar{T}_{n, {\rm tot}}^{(n)}= \left(\frac{C_1}{xy-1}+C_2\right)[xy(x+y)]^{n-1}-\frac{C_1}{xy-1}(x+y)^{n-1}\,,
\end{equation}
in which $C_1$ and $C_2$ are two constants with $C_1=\frac{x(x^2-1)+y(y^2-1)}{6(x+y)}(x+y-2xy)$ and $C_2=\bar{T}_{1, {\rm tot}}^{(1)}=\frac{(x+y)^3-(x+y)}{6}-xy$.

Inserting Eq.~(\ref{eq:TXxy}) into Eq.~(\ref{eq:Ttotxy}) yields
\begin{equation}\label{eq:T2xy}
\begin{aligned}
T_{n, {\rm tot}}^{(n)} = xy\,T_{n-1, {\rm tot}}^{(n-1)}+
\left(\frac{C_1}{xy-1}+C_2\right)[xy(x+y)]^{n-1}-\,\\
\frac{C_1}{xy-1}(x+y)^{n-1} \,.
\end{aligned}
\end{equation}
Considering $T_{1, {\rm tot}}^{(1)} = \frac{(x+y)^3-(x+y)}{6}$, Eq.~(\ref{eq:T2xy}) is resolved by induction to obtain
\begin{equation}\label{eq:T2xy2}
\begin{aligned}
T_{n, {\rm tot}}^{(n)} = &\frac{D_2}{xy-x-y}(x+y)^n+ \,\\
&\left(\frac{D_3}{xy}-\frac{D_2(x+y)}{xy(xy-x-y)}-\frac{D_1(x+y)}{xy(x+y-1)}\right)(xy)^n \,\\
&+\frac{D_1}{xy(x+y-1)}[xy(x+y)]^n \,.
\end{aligned}
\end{equation}
where $D_1=\frac{C_1}{xy-1}+C_2$, $D_2=\frac{C_1}{xy-1}$, and $D_3=T_{1, {\rm tot}}^{(1)}=\frac{(x+y)^3-(x+y)}{6}$.

From Eq.~(\ref{Nn}), we have
$(x+y)^n=\frac{x+y-1}{x+y-2}N_n-\frac{x+y}{x+y-2}$ and
$n=\log_{x+y}\left(\frac{x+y-1}{x+y-2}N_n-\frac{x+y}{x+y-2}\right)$. In addition, we assume that
$G_1=\frac{D_2}{xy-x-y}$, $G_2=\frac{D_3}{xy}-\frac{D_2(x+y)}{xy(xy-x-y)}-\frac{D_1(x+y)}{xy(x+y-1)}$,
and $G_3=\frac{D_1}{xy(x+y-1)}.$
Then, Eq.~(\ref{eq:T2xy2}) can be recast as a function of network order $N_n$
\begin{eqnarray}\label{Totalxy}
T_{n, {\rm tot}}^{(n)}&=&
G_1\left(\frac{x+y-1}{x+y-2}N_n-\frac{x+y}{x+y-2}\right) \,\nonumber\\
&\quad&+G_2\left(\frac{x+y-1}{x+y-2}N_n-\frac{x+y}{x+y-2}\right)^{\frac{\ln(xy)}{\ln(x+y)}} \,\nonumber\\
&\quad&+G_3\left(\frac{x+y-1}{x+y-2}N_n-\frac{x+y}{x+y-2}\right)^{\frac{\ln[xy(x+y)]}{\ln(x+y)}} \,.
\end{eqnarray}
Thus, by definition given by Eq.~(\ref{ATT}), the average trapping time $\langle T \rangle_n$ is
\begin{eqnarray}\label{MFPTxyTn}
\langle T \rangle_n&=&
\frac{T_{n, {\rm tot}}^{(n)}}{N_n-1}=
\frac{G_1}{N_n-1}\left(\frac{x+y-1}{x+y-2}N_n-\frac{x+y}{x+y-2}\right) \,\nonumber\\
&\quad&+\frac{G_2}{N_n-1}\left(\frac{x+y-1}{x+y-2}N_n-\frac{x+y}{x+y-2}\right)^{\frac{\ln(xy)}{\ln(x+y)}} \,\nonumber\\
&\quad&+\frac{G_3}{N_n-1}\left(\frac{x+y-1}{x+y-2}N_n-\frac{x+y}{x+y-2}\right)^{\frac{\ln[xy(x+y)]}{\ln(x+y)}}\,.
\end{eqnarray}
In the limit of large network size, $\langle T \rangle_n$ is approximatively equal to
\begin{equation}\label{MFPTxy}
\langle T \rangle_n \sim
(N_n)^{\ln xy/\ln (x+y)}=(N_n)^{\theta(x,y)}\,.
\end{equation}

Equation~(\ref{MFPTxy}) shows that, the dominating scaling of average trapping time
grows as a power-law function of network size $N_n$ with the exponent $\theta(x,y)=\ln xy/\ln
(x+y)$ being compatible with the bounds derived in~\cite{TeBeVo09}. Note that a scaling similar to  $\theta(x,y)$ was previously obtained for general fractals without the scale-free phenomenon~\cite{Bobe05}. It is also worth emphasizing that for some limiting $x$ and $y$, $\theta(x,y)$ recovers some previous results. For $x=1$ and $y=2$, $\theta(x,y)$ reduces to $\ln 2/\ln
3$ previously obtained in~\cite{ZhQiZhXiGu09}; for $x=1$ and $y=3$, $\theta(x,y)$ is reduced to $\ln 3/\ln 4$~\cite{ZhXiZhLiGu09}; while for $x=2$ and $y=2$, $\theta(x,y)$ corresponds to 1~\cite{ZhXiZhLiGu09}. The consistency confirms that Eq.~(\ref{MFPTxy}) is valid.

\subsubsection*{Result analysis}

As mentioned above, the average trapping time $\langle T \rangle_n$ is a highly desirable quantity characterizing the trapping problem in a network. For trapping issue on the $(x,y)-$flowers, $\langle T \rangle_n$ is dominated by the exponent $\theta(x,y)={\frac{\ln xy}{\ln
(x+y)}}$. It is not difficult to verify that for different values of $x+y$, the exponent $\theta(x,y)$ increases with the sum of $x$ and $y$.
Hence, $x+y$ determines the trapping efficiency for random walks on the $(x,y)-$flowers:
the larger the sum $x+y$, the less the efficiency. This can be understood from the following arguments. When $x+y$ become larger, the $(x,y)-$flowers are more homogeneous, since the exponent $\gamma$ of the power-law degree distribution grows with increasing $x+y$. According to the previous result that the trapping efficiency in scale-free networks decreases when the exponent $\gamma$ increases~\cite{KiCaHaAr08}, we can know why the sum $x+y$ influences strongly the leading behavior of the average trapping time.

For the case when $x+y$ is fixed, Eq.~(\ref{MFPTxy}) shows that $\langle T \rangle_n$ is determined by the difference between $x$ and $y$ (since we have assumed that $x\leq y$): the smaller the difference $y-x$, the larger the value $\langle T \rangle_n$. As shown in section~\ref{SecModel}, for the $(x,y)$-flowers with fixed $x+y$, they have the same degree sequence and thus the same degree distribution. Thus, we can conclude that power-law degree distribution alone is insufficient to determine the trapping efficiency in scale-free networks.

As pointed out above, the power-law degree distribution alone cannot determine the behavior of trapping problem in scale-free networks. Then, it is natural and important to ask: which structural property plays a crucial role in the trapping efficiency in the $(x,y)$-flowers with the same value of $x+y$? Since for those $(x,y)$-flowers with an identical sum of $x+y$, they have the same average node degree, the same degree distribution, and the same (zero) clustering coefficient; moreover, for the family of $(x,y)$-flowers at the same generation, their initial nodes have an identical degree, we argue that the difference for the average trapping time shown in Eq.~(\ref{MFPTxy}) is only due to the fractality. For those $(x,y)-$flowers having the same parameter of $x+y$, when $x$ increases from 1 to its maximum, their fractal dimension $d_{\rm f}=\frac{\ln (x+y)}{\ln x}$ decreases from $\infty$ to its minimum, while the exponent $\theta(x,y)$ increases from its minimum to maximum. Concretely, the trapping efficiency in the $(x,y)-$flowers with the same $x+y$ is dominated by the fractal dimension: the larger the fractal dimension, the more efficient the trapping process. Particularly, when $x=1$, the network has an infinite fractal dimension, which corresponds to the most efficient structure for trapping, among all $(x,y)-$flowers with the same sum of $x$ and $y$.

It is worthy of mentioning that the impact of fractal dimension on the trapping efficiency can also be seen by comparing the average trapping time in the $(1,2)-$flower~\cite{ZhQiZhXiGu09} and the two-dimensional Sierpinski gasket~\cite{KaBa02PRE}. The trapping process is more efficient in the $(1,2)-$flower than in the two-dimensional Sierpinski gasket, in spite of the fact that both networks have the same numbers of nodes and edges at any iteration. The difference for average trapping time in the two networks is at least partially attributed to the fractality. The $(1,2)-$flower is non-fractal with an infinite fractal dimension, while the corresponding Sierpinski gasket has a fractal dimension of $1+\frac{\ln 2}{\ln 3}$.

\subsection{Average trapping time for hybrid networks}

We proceed to show that the above relation between average trapping time and fractal dimension for trapping problem in the deterministic scale-free $(x,y)$-flowers are also true for the random hybrid networks $H_n$.

Similar to the $(x,y)$-flowers, we first establish the relation dominating the evolution for $T_i^{(n)}$ with iteration $n$. Let's consider a node $i$ in $H_n$. By construction, we know that, upon the growth of the networks to iteration $n+1$, the degree of node $i$ increases from $k_i(n)$ to $k_i(n+1)=2\,k_i(n)$. Among these $k_i(n+1)$ neighbors, some are created at iteration $n+1$, each of which is either generated by the first iterative method with probability $q$, or created by the second iterative method with probability $1-q$. We now consider an unbiased random walk in $H_{n+1}$: let $A$ denote the FPT for going from node $i$ to any of its $k_i(n)$  old neighbors, i.e., those nodes directly connected to $i$ at iteration $n$; let $B$ stand for the FPT from a new neighbor of $i$, which is generated through the first iterative method, to one of its $k_i(n)$  old neighbors; let $C$ represent the FPT from a new node created at iteration $n+1$, emerging simultaneously with a new neighbor of $i$ that was generated by the first iterative method and is attached to this new neighbor of $i$, to an old neighbor of $i$; and let $D$ be the FPT originating at any of $i'$s new neighbors created by the second iterative method to one of $i'$s old neighbors. Then the following backward equations can be established:
\begin{eqnarray}\label{MFPT1}
\left\{
\begin{array}{rcl}
A&=&q\left[\frac{1}{2}+\frac{1}{2}(1+B)\right]+(1-q)(1+D)\,, \\
B&=&\frac{1}{2}(1+A)+\frac{1}{2}(1+C) \,,\\
C&=&\frac{1}{2}+\frac{1}{2}(1+B) \,,\\
D&=&\frac{1}{2}+\frac{1}{2}(1+A) \,,\\
\end{array}
\right.
\end{eqnarray}
which leads to $A = \frac{{12}}{{q + 3}}$. Thus, we can obtain an important relation $T_i^{(n+1)}=\frac{{12}}{{q + 3}}\,T_i^{(n)}$.

Having obtaining the evolution rule of trapping time when the hybrid networks grow, we now compute the the average trapping time $\langle T \rangle_n$. For simplicity, In the following text, we will use the same notations as those for the $(x,y)-$flowers defined above. Analogous to
$(x,y)-$flowers, we have the following equation:
\begin{equation}\label{Ttot}
T_{n, {\rm tot}}^{(n)} = T_{n - 1, {\rm tot}}^{(n)} +
\bar{T}_{n, {\rm tot}}^{(n)}= \frac{{12}}{{q + 3}}\,T_{n - 1, {\rm tot}}^{(n-1)} + \bar{T}_{n, {\rm tot}}^{(n)}\,.
\end{equation}
Hence, the problem of determining $T_{n, {\rm tot}}^{(n)}$ is reduced to evaluating $\bar{T}_{n, {\rm tot}}^{(n)}$ that can be obtained as follow.

According to the iterative rule, at a given generation, for each edge connecting two nodes $i$ and $j$ in $H_n$, it will generate two
new nodes at the subsequent iteration $n+1$. Both nodes are either generated by the first iterative method with probability $q$ and labeled by $w_1$ and $w_2$, or generated by the second iterative method with probability $1-q$ and denoted by $w'_1$ and $w'_2$. For $w_1$ and $w_2$, their trapping times obey the relations:
\begin{eqnarray}\label{oldnew1_1}
\left\{
\begin{array}{rcl}
T_{w_1}^{(n+1)}$=$\frac{1}{2}[1+T_{w_2}^{(n+1)}]+\frac{1}{2}[1+T_{i}^{(n+1)}] \,,\\
T_{w_2}^{(n+1)}$=$\frac{1}{2}[1+T_{w_1}^{(n+1)}]+\frac{1}{2}[1+T_{j}^{(n+1)}] \,.\\
\end{array}
\right.
\end{eqnarray}
From Eq.~(\ref{oldnew1_1}), we obtain
\begin{equation}\label{oldnew1_2}
T_{w_1}^{(n+1)}+T_{w_2}^{(n+1)}=4+T_{i}^{(n+1)}+T_{j}^{(n+1)}\,.
\end{equation}
Similarly, for $w'_1$ and $w'_2$, we have
\begin{equation}\label{oldnew2_1}
T_{w'_1}^{(n+1)}=T_{w'_2}^{(n+1)}=1+\frac{1}{2}\left(T_{i}^{(n+1)}+T_{j}^{(n+1)}\right) \,,
\end{equation}
yielding
\begin{equation}\label{oldnew2_2}
T_{w'_1}^{(n+1)}+T_{w'_2}^{(n+1)}=2+T_{i}^{(n+1)}+T_{j}^{(n+1)}\,.
\end{equation}

Summing Eq.~(\ref{oldnew1_2}) and Eq.~(\ref{oldnew2_2}) over all the $M_n$ edges in $H_n$ and considering the iterating probability for each preexisting edge at the generation $n$, we get
\begin{eqnarray}\label{totrec1}
\bar T _{n + 1,{\rm tot}}^{(n + 1)} &=&(2q + 2){M_n} + \sum \limits_{i \in {\Omega_n}} {\left({k_i}(n) \times T_i^{(n + 1)}\right)} \nonumber \\
 &=& (2q + 2){4^n} + 2\bar T_{n,{\rm tot}}^{(n + 1)} + {2^2}\bar T_{n - 1,{\rm tot}}^{(n + 1)} +  \cdots \nonumber \\
   &\quad&+ {2^n}\bar T_{1,{\rm tot}}^{(n + 1)} + {2^n}\bar T_{0,{\rm tot}}^{(n + 1)} \,.
\end{eqnarray}
In addition, from Eq.~(\ref{totrec1}), we can write out
$\bar{T}_{n+2, {\rm tot}}^{(n+2)}$ as
\begin{eqnarray}\label{totrec3}
\bar T _{n + 2,{\rm tot}}^{(n + 2)} &=& (2q + 2){4^{n + 1}} + 2\bar T_{n + 1,{\rm tot}}^{(n + 2)} + {2^2}\bar T_{n,{\rm tot}}^{(n + 2)} +  \cdots  \nonumber \\
&\quad&+ {2^{n + 1}}\bar T_{1,{\rm tot}}^{(n + 2)} + {2^{n + 1}}\bar T_{0,{\rm tot}}^{(n + 2)} \,.
\end{eqnarray}

Equation~(\ref{totrec3}) minus Eq.~(\ref{totrec1}) times
$\frac{24}{q+3}$ and
applying the relation of $T_i^{(n+2)}=\frac{{12}}{{q + 3}}\,T_i^{(n+1)}$, we obtain the
following recursive relation:
\begin{equation}\label{totrec4}
\bar T _{n + 2,{\rm tot}}^{(n + 2)} = \frac{{48}}{{q + 3}}\bar T _{n + 1,{\rm tot}}^{(n + 1)}- \frac{{ 6+ 4q-2{q^2}}}{{q + 3}} \times {4^{n + 1}} \,.
\end{equation}
With the initial value of $\bar T _{1,{\rm tot}}^{(1)}=6+q$, Eq.~(\ref{totrec4})
is solved inductively
\begin{equation}\label{totrec5}
\bar T _{n,{\rm tot}}^{(n)} = \frac{48+q^2-q}{9-q}\left(\frac{48}{3+q}\right)^{n-1}+\frac{{(q + 1)(6 - 2q)}}{{9 - q}}{4^{n - 1}}\,.
\end{equation}

Plugging Eq.~(\ref{totrec5}) into Eq.~(\ref{Ttot}) yields
\begin{eqnarray}\label{totrec6}
T_{n,{\rm tot}}^{(n)} &=& \frac{{12}}{{q + 3}}T_{n - 1,{\rm tot}}^{(n - 1)} + \frac{48+q^2-q}{9-q}\left(\frac{48}{3+q}\right)^{n-1} \nonumber \\
&\quad&+\frac{{(q + 1)(6 - 2q)}}{{9 - q}}{4^{n - 1}} \,.
\end{eqnarray}
Considering $T_{1, \rm tot}^{(1)} = 10$,
Eq.~(\ref{totrec6}) can be resolved by induction.
When $q=0$,
\begin{eqnarray}\label{totrec7_1}
T_{n,\rm tot}^{(n)}&=& \frac{{{4^n}}}{{18}}(8 \times {4^n} + 3n + 10);
\end{eqnarray}
when $0<q \leq 1$,
\begin{eqnarray}\label{totrec7_2}
T_{n,{\rm tot}}^{(n)}& =&\frac{{(q + 3)(48+q^2-q)}}{{36( 9 - q)}}\left(\frac{{48}}{{q + 3}}\right)^n  \nonumber \\
&\quad&+\frac{{(3-q)(q + 3)(q + 1)}}{{2q(9-q)}}{4^n}\nonumber \\
&\quad&- \frac{{(q + 3)({q^2} - q + 3)}}{{18q}}\left(\frac{{12}}{{3 + q}}\right)^n \,.
\end{eqnarray}

Using Eq.~(\ref{ATT}), we can find the analytical solution for average trapping time in $H_n$, from which we can obtain the exact dependence relation of $\langle T \rangle_n$ on the network size $N_n$. For $q=0$,
\begin{equation}\label{MFPT9}
 \langle T
\rangle_n=\frac{3N_n-4}{36(N_n-1)}\left[12N_n+ \frac{3\,\ln
\left(\frac{3}{2}N_n-2\right)}{2\,\ln 2}-6\right];
\end{equation}
while for $0<q \leq 1$
\begin{eqnarray}\label{MFPT10}
{\left\langle T \right\rangle _n} &=&  \frac{1}{{{N_n} - 1}}\frac{{(q + 3)(48+q^2 - q)}}{{36(9 - q)}}\left(\frac{3}{2}N_n - 2\right)^{\frac{\ln \frac{48}{q + 3}}{\ln 4}} \nonumber \\
&\quad&+\frac{1}{{{N_n} - 1}}\frac{{(3-q)(q + 3)(q + 1)}}{{2q( 9 - q)}}\left(\frac{3}{2}N_n-2\right)\nonumber \\
&\quad&- \frac{1}{{{N_n} - 1}}\frac{{(q + 3)({q^2} - q + 3)}}{{18q}}\left(\frac{3}{2}{N_n} - 2\right)^{\frac{\ln \frac{12}{q + 3}}{\ln 4}}.\nonumber \\
\end{eqnarray}

\begin{figure}
\begin{center}
\includegraphics[width=0.9\linewidth,trim=0 30 0 25]{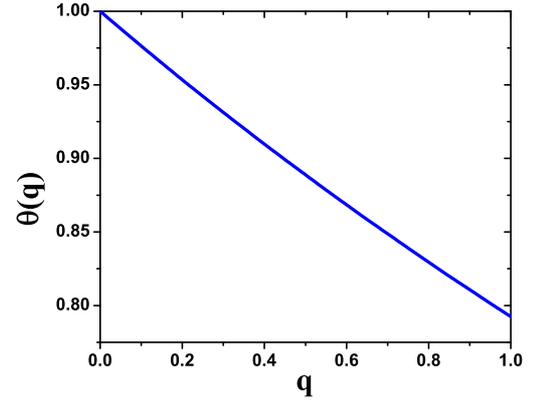}
\caption{(Color online) Relation between $\theta(q)$ and $q$.} \label{thetaq}
\end{center}
\end{figure}

From Eqs.~(\ref{MFPT9}) and~(\ref{MFPT10}) we can easily see that in the full range of $0 \leq q \leq 1$, the average trapping time
$\langle T \rangle_n$ in the hybrid networks behaves as a power-law function of
network order $N_n$, with the exponent $\theta(q)=\ln \frac{12}{q+3}/ \ln 4$ decreasing with parameter
$q$. Concretely, when $q$ grows from 0 to 1, $\theta(q)$ drops from 1 to $\ln 3/ \ln 4$, see Fig.~\ref{thetaq}. In addition, when $q$ enhances from 0 to 1, the fractal dimension $d_{\rm f}$ of the hybrid networks increases from 2 to $\infty$. Thus, in the random hybrid networks, the diffusion efficiency also enhances with the increasing fractal dimension as that found for the $(x,y)-$flowers.

\section{conclusions}

We have studied the trapping problem in two families of networks, i.e., the $(x,y)-$flowers and random hybrid networks. The former is deterministic, while the latter is stochastic. Both display rich and specific structural properties and can recover some prominent features of various real networked systems. We focused on a particular case of trapping issue with the immobile trap located at either of the initial nodes of the networks. We derived analytically the average trapping time and demonstrated that they grow as a power-law function of the network size with the exponent varying with network parameters. We showed that the power-law degree distribution (even degree sequence) itself is not sufficient to determine the behavior of average trapping time of random walks on fractal scale-free networks. Instead, we presented that fractal dimension plays a predominant role in the scaling of average trapping in both families of fractal scale-free networks under consideration. Finally, it should be stressed that here we only studied particular networks with the trap fixed on a specific node having the highest degree, whether the conclusion also holds for a general fractal scale-free network or for a non-hub node needs further research in the future.

\section*{Acknowledgment}

We thank Yuan Lin for preparing this manuscript. This work was supported by the National Natural Science Foundation of China under Grant No. 61074119.


\end{document}